\documentclass[12pt,preprint]{article}

\usepackage{amssymb}
\usepackage{amsmath}
\usepackage{graphics}
\usepackage{epsfig}
\renewcommand{\vec}[1]{{\bf #1}}
\newcommand{\eqb}{\begin{equation}}
\newcommand{\eqe}{\end{equation}}
\newcommand{\dmb}{\begin{displaymath}}
\newcommand{\dme}{\end{displaymath}}
\newcommand{\pd}{\partial}

\newcommand{\eab}{\begin{eqnarray}}
\newcommand{\eae}{\end{eqnarray}}
\newcommand{\ra}{\right\rangle}
\newcommand{\la}{\left\langle}
\newcommand{\e}{\mbox{e}}
\newcommand{\be}{\begin{equation}}
\newcommand{\ee}{\end{equation}}

\begin{document}
\begin{titlepage}

\begin{center}
\Large{Charge-density waves in deconfining SU(2) Yang-Mills thermodynamics }\vspace{1.5cm}\\ 
\large{Carlos Falquez$^*$, Ralf Hofmann$^{**}$, and Tilo Baumbach$^*$}
\end{center}
\vspace{2.0cm} 
\begin{center}
{\em $\mbox{}^{*}$ Laboratorium f\"ur Applikationen der Synchrotronstrahlung (LAS)\\ 
Karlsruher Institut f\"ur Technologie (KIT)\\  
Postfach 6980\\ 
76128 Karlsruhe, Germany}
\end{center}
\vspace{1.0cm} 
\begin{center}
{\em $\mbox{}^{**}$ Institut f\"ur Theoretische Physik\\ 
Universit\"at Heidelberg\\ 
Philosophenweg 16\\ 
69120 Heidelberg, Germany}
\end{center}
\vspace{1.0cm}
\begin{abstract}
At one-loop accuracy we compute, characterize, and discuss 
the dispersion laws for the three low-momentum branches of propagating 
longitudinal, electric U(1) fields in the effective theory for 
the deconfining phase of pure SU(2) Yang-Mills thermodynamics. With an electric-magnetically dual 
interpretation of SU(2)$_{\mbox{\tiny CMB}}$ we argue 
that upon a breaking of plasma isotropy and 
homogeneity, introduced e.g. by a temperature gradient, the longitudinal 
modes could conspire to provide magnetic seed fields for 
magneto-hydrodynamical dynamos inside structures of galaxy, galaxy-cluster, and cosmological scales. 
Such a scenario ultimately links structure with seed-field formation. 
As judged from the present cosmological epoch, the maximally available coherent field strength of 10$^{-8}$\,Gauss from 
SU(2)$_{\mbox{\tiny CMB}}$ matches with the upper bound 
for cosmological present-day field strength derived from 
small-angle anisotropies of the cosmic microwave background.      
 \end{abstract} 
\end{titlepage}

\section{Introduction}

In its two phases with propagating gauge fields the thermal ground states of 
SU(2) Yang-Mills thermodynamics exhibit macroscopic behavior which in some repects resembles that 
of metals. By a selfconsistent spatial coarse-graining in Euclidean signature over quantum fluctuations of trivial 
\cite{'t Hooft Veltman,Zinn-Justin} and nontrivial topology \cite{HarringtonShepard,Nahm,GrossPisarskiYaffe,LeeLu,KraanVanBaal,Diakonov} 
precise estimates of the ground-state physics are obtained in both phases \cite{HerbstHofmann2004,Hofmann2005,Hofmann2007}. 
The thermal ground states determine the properties of effective and very weakly or noninteracting thermal quasiparticle 
excitations \cite{HerbstHofmann2004,Hofmann2005,Hofmann2006,Hofmann2007}. While in the deconfining phase one direction of the 
three dimensional Lie algebra of SU(2) remains massless ('photon') the 
only (noninteracting) propagating gauge mode in the preconfining phase is massive due to the  
dual Meissner effect and the decoupling of massive modes at the deconfining-preconfining phase boundary. In the deconfining phase 
nontrivial dispersion 
laws for the three polarizations states of the massless U(1) gauge mode occur through resummed one-loop radiative 
corrections. In \cite{LudescherHofmann2008} the dispersion for transversely polarized `photons' 
was computed and further characterized in 
\cite{FalquezHofmannBaumbach2010}. In the context of SU(2)$_{\mbox{\tiny CMB}}$ \cite{Hofmann2005,GiacosaHofmann2005} 
this\footnote{This postulate states that, fundamentally, the propagation of 
electromagnetic waves is described by an SU(2) rather than a U(1) gauge principle. Observations of the Cosmic Microwave Background (CMB)
then suggest \cite{Arcade2,Schwarz2010} that the critical temperature for the deconfining-preconfining phase 
transition coincides with the CMB's present temperature \cite{SzopaHofmann2007,Hofmann2009}, hence the term 
SU(2)$_{\mbox{\tiny CMB}}$. Interplaying with an axion field of Planckian origin 
\cite{GiacosaHofmann2005,Sikivie,Wilczek,Frieman} the thermal-ground state physics of 
SU(2)$_{\mbox{\tiny CMB}}$ could be responsible for the (dominant) 
dark-energy component in today's mix of cosmological fluids.} dispersion predicts a gap in the 
spectrum of black-body radiation at low temperatures and frequencies, see \cite{FalquezDA} 
for an investigation of experimental signatures. The dispersion of the longitudinal polarization has not been 
investigated so far. Longitudinal modes are absent on tree level. Moreover, according to an electric-magnetically dual 
interpretation of SU(2)$_{\mbox{\tiny CMB}}$ they represent longitudinally propagating, long-wavelength magnetic 
fields whose intensity can not be measured by a detector.  

In the present work we investigate the propagation properties of these longitudinal modes. 
In deconfining SU(2) Yang-Mills thermodynamics (no electric-magnetically dual 
interpretation) they are associated with the propagation of radiatively induced, electric charge densities. 
The computation is performed in analogy to \cite{LudescherHofmann2008}, where the propagation of transverse modes was studied, 
by a one-loop selfconsistent resummation of the longitudinal component of the massless mode's 
polarization tensor. As a result and in contrast to the transverse case, three branches of longitudinal modes occur 
at low momenta. Scaling out temperature, their high-temperature dispersion is temperature independent. One branch resembles a finite-support 
lightlike dispersion. Micoscopically, such a behavior should be associated with the slow motion of 
stable magnetic monopoles and antimonopoles released by large-holonomy (anti)calorons upon dissociation \cite{Diakonov,LKGH2008}. 
The other branches exhibit superluminal group velocities which suggest their association 
with instantaneous spatial correlations between short-lived but fast-moving magnetic 
monopoles and antimonopoles. (Anti)monopoles of this type occur in small-holonomy (anti)calorons \cite{Diakonov,LKGH2008}.        

The paper is organized as follows. In Sec.\,\ref{setupcomp} we explain how the temporal component 
of the polarization tensor of the massless mode is related to the screening function $F$ in the dispersion law and how $F$ is computed selfconsistently at one-loop accuracy in the 
deconfining phase of SU(2) Yang-Mills thermodynamics. Sec.\,\ref{Nec} presents numerical 
results for the three branches that occur at low momenta. We also present fits to affine power 
laws of the temperature dependence of characteristic points in the dispersion laws. The energy density of each 
branch is computed for the high-temperature regime in Sec.\,\ref{htened}, and a connection to cosmological magnetic seed 
fields inducing inside astrophysical structures regular magnetic fields, which were 
amplified by magneto-hydrodynamical (MHD) dynamos, is made. In Sec.\,\ref{sum} we summarize our results.

\section{Computation of dispersion laws for longitudinal modes\label{setupcomp}}

This section provides the theoretical background on the physics of longitudinal U(1) modes in the 
sector of tree-level massless gauge modes in deconfining SU(2) Yang-Mills thermodynamics \cite{Hofmann2005}. 
Greek (latin) indices refer to spacetime (space) coordinates. All calculations are performed in 
physical unitary-Coulomb gauge where the former gauge condition relates to a choice of 
direction in the Lie algebra for the caloron-anticaloron induced, inert and adjoint scalar 
field $\phi$ \cite{HerbstHofmann2004}. For definiteness one demands 
$\phi^a\equiv 2|\phi|\delta^{a3}$ where $|\phi|\equiv\sqrt{\frac12\mbox{tr}\,\phi^2}$. 
The Coulomb condition $\pd_i a^3_i=0$ requires the modes of the massless, effective gauge 
field $a^3_\mu$ to be spatially transverse. Given canonical behavior 
at spatial infinity, this fixes the intact U(1) gauge symmetry physically in a unique way. (The freedom of applying time-dependent 
U(1) gauge transformations is lost by demanding $a^3_0$ to decay to zero towards spatial infinity.)     

\subsection{Preliminaries}

\begin{figure}
\begin{center}
\leavevmode
\leavevmode
\vspace{4.9cm}
\includegraphics{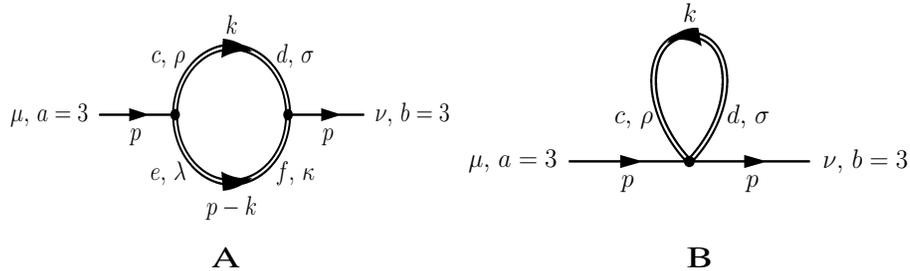}
\end{center}
\caption{\protect{\label{Fig-1}} Two one-loop diagrams which potentially contribute to the 
polarization tensor of the massless mode of deconfining SU(2) Yang-Mills thermodynamics. Single lines are associated with the external 'photon' field, 
double lines depict the propagation of tree-level heavy modes in the thermal loops. (These latter modes occur on their mass-shell 
only, $k^2=m^2=4\,e^2|\phi|^2$, where $e$ denotes the effective gauge coupling \cite{Hofmann2005,Hofmann2007}.)}      
\end{figure}

Let us briefly review how the calculation of the one-loop dispersions of the longitudinal 
polarization state of the massless mode proceeds in the effective theory for the deconfining phase 
\cite{Hofmann2005,Hofmann2007,LudescherHofmann2008,SHG2006}. All calculations are performed in 
unitary-Coulomb gauge. In this physical gauge the formulation of constraints on loop-momenta 
is simple \cite{Hofmann2006}. As in the transverse case our strategy is to compute in Minkowskian signature the dispersion due to 
diagram B of Fig.\,\ref{Fig-1} and to subsequently check selfconsistency by demonstrating that diagram A is 
nil on the so-determined shell $p^2=F$ where $F$ denotes the longitudinal screening function, see below.     

On the one-loop level\footnote{In applications (SU(2)$_{\mbox{\tiny CMB}}$) 
this accuracy is sufficient owing to the fast covergence of the effective loop expansion \cite{Hofmann2006,SHG2006,KH2007}.} 
the task is the determination of the invariant $F$ in the 00 component of the interacting (imaginary-time) propagator (for an extended 
discussion see Sec.\,3.5.1 of \cite{FalquezDA}) 
\eqb
\label{propTLMWWrep}
D^{\mbox{\tiny TLM}}_{ab,\mu\nu}(p) = 
-\delta_{a3}\delta_{b3}\Big \lbrace P^T_{\mu\nu}\frac{1}{p^2+G}+\frac{p^2}{\vec{p}^2}\frac{1}{p^2+F}\,u_\mu u_\nu\Big \rbrace\,,
\eqe
where 
\begin{align}
P^T_{00}& = P^T_{0i} = P^T_{i0} = 0\,,\\
P^T_{ij}& = \delta_{ij} - \frac{p_{i}p_{j}}{\textbf{p}^2}\,,
\end{align}
and $u_\mu=\delta_{\mu 4}=\delta_{\mu 0}$ represents the four-velocity of the
heat bath. Due to the intact U(1) gauge symmetry the polarization tensor $\Pi_{\mu\nu}$ of 
the massless mode is 4D transverse,
\eqb
\label{transPi}
p_\mu \Pi_{\mu\nu}=0\,,
\eqe
and the following decomposition holds
\eqb
\label{Pidec}
\Pi_{\mu\nu}=G(p_4,\vec{p})\,P^T_{\mu\nu}+F(p_4,\vec{p})\,P^L_{\mu\nu}\,,
\eqe
where 
\eqb
\label{PL}
P^L_{\mu\nu}\equiv\delta_{\mu\nu}-\frac{p_\mu p_\nu}{p^2}-P^T_{\mu\nu}\,.
\eqe
Obviously, $P^L_{ij}$ is spatially longitudinal and, according to the decomposition in Eq.\,(\ref{Pidec}),  
$\Pi_{44}=F\left(1-\frac{p_4^2}{p^2}\right)$. In Eq.\,(\ref{Pidec}) we suppress the temperature dependences of the invariants 
$G$ and $F$ arising by a resummation of one- to infinite-fold insertions of $\Pi_{\mu\nu}$ into the 
free propagator $D^{\mbox{\tiny TLM},0}_{ab,\mu\nu}(p)$ given as 
\eqb
\label{propTLMWWrepfree}
D^{\mbox{\tiny TLM},0}_{ab,\mu\nu}(p) = 
-\delta_{a3}\delta_{b3}\Big\lbrace\frac{P^T_{\mu\nu}}{p^2}+\frac{u_\mu u_\nu}{\vec{p}^2}\Big\rbrace\,.
\eqe
For Minkowskian signature it was argued in \cite{SHG2006} 
that in both limits $\vec{p}\to 0$ at $p_0=0$ and 
$|p_0|\to |\vec{p}|$ the behavior of function $F$ is not interesting. 
Namely, we have that $|F|\to\infty$ in the former and 
$F\to 0$ in the latter case. While $|F|\to\infty$ decouples the longitudinal mode from 
the spectrum of propagating excitations $F\to 0$ clashes with the free limit 
where, according to Eq.\,(\ref{propTLMWWrepfree}), the longitudinal mode is instantaneous. 
(Recall that $|p_0|\to |\vec{p}|$ or $p^2=0$ was assumed.) As we will show below, 
this contradiction is resolved by the selfconsistent determination of the 
dispersion law: With a cutoff on $|\vec{p}|$ the situation $|p_0|\to |\vec{p}|$ 
takes place as the high-temperature limit in one of three possible 
branches.  

In Fig.\,\ref{Fig-1} the two diagrams, which potentially contribute to $\Pi_{\mu\nu}$, are depicted.  
In Minkowskian signature restriction to the shell $p^2=F$ yields \cite{KapustaGale}
\eab
\label{P00Fimpl}
F(p_4,\vec{p})&=&\left(1-\frac{p_4^2}{p^2}\right)^{-1}\,\Pi_{44}\ \to \ F(p_0,\vec{p})=-\left(1-\frac{p_0^2}{p^2}\right)^{-1}\,\Pi_{00}\ \Rightarrow\nonumber\\ 
\vec{p}^2&=&\Pi_{00}\,.
\eae
The right-hand side of the gap equation $\vec{p}^2=\Pi_{00}$ implicitly depends on $F$ through a vertex constraint that must be applied to diagram B in 
the effective theory \cite{Hofmann2005}, and for a given value\footnote{Due to the isotropy of the thermal 
plasma the dependence of $\Pi^{00}$ on $\vec{p}$ is through $|\vec{p}|$ only.} of $|\vec{p}|$ the value of $F$ 
is determined by numerical inversion of the last-line equation in (\ref{P00Fimpl}). Finally, 
the propagation of the longitudinal mode is determined by  
\eab
\label{displaw}
\omega^2(\vec{p})&=&\vec{p}^2+\mbox{Re}\,F(\vec{p})\,,\nonumber\\ 
\gamma(\vec{p})&=&-\frac{1}{2\omega}\mbox{Im}\,F(\vec{p})\,,
\eae
where $\omega$ and $\gamma^{-1}(\vec{p})$ represent energy and lifetime of the longitudinal 
mode at momentum $\vec{p}$. A finite lifetime or an imaginary contribution to $F$ can, however, only be generated by diagram A (`photon' decay into two massive vector modes). 
However, diagram A turns out to vanish at $p^2=F$ with $F$ determined by diagram B only, 
see also \cite{LudescherHofmann2008} for the transverse case.

\subsection{Gap equation}

Here we give the equation explicitly that determines the values of $F$ at a given momentum modulus $|\vec{p}|$. 
It is advantageous to express energy $\omega$ and $|\vec{p}|$ in terms of the dimensionless ratios 
$X\equiv\frac{|\vec{p}|}{T}$ and $Y\equiv\frac{\omega}{T}$ where $T$ denotes temperature. The gap equation $\vec{p}^2=\Pi_{00}$ then is 
re-cast as
\begin{align}
X^2=\frac{\Pi_{00}}{T^2}\,.
\label{P00Fonshell}
\end{align}
Without restriction of generality we may assume that $\vec{p}=|\vec{p}|e_z$ where $e_z$ is the unit vector in 3-direction. 
After a change to cylindrical coordinates and a re-scaling of the integration variables the contribution from 
diagram B of Fig.\,\ref{Fig-1} to the right-hand side of Eq.\,(\ref{P00Fonshell}) reads \cite{SHG2006}
\begin{align}
  X^2 \, &= \, \frac{\Pi_{00}}{T^2} \notag \\
         &= \, \int \mathrm d \xi\,\int \mathrm d\rho\,\,  2e^2\lambda^{-3}
                \left(3+\frac{\rho^2+\xi^2}{4e^2}\right)\,\rho\,
                        \frac{n_B\left(2\pi \lambda^{-3/2}\sqrt{\rho^2+\xi^2+4e^2}\right)}{\sqrt{\rho^2+\xi^2+4e^2}}
 \\
                        &\equiv \, \int \mathrm d \xi\,\int \mathrm d\rho\,\, h_F \left( \xi,\rho, \lambda \right)
                        \,,
  \label{defF}
\end{align}
where $n_B(x)\equiv \frac{1}{\e^x-1}$, $\lambda\equiv\frac{2\pi T}{\Lambda}$, and $\Lambda$ denotes the Yang-Mills scale. 
The temperature dependence of the effective gauge coupling $e$ is a consequence of thermodynamical consistency of the pressure at the one-loop 
level, and $e$ exhibits a logarithmic pole at $T_c=\frac{\Lambda}{2\pi}\lambda_c$  ($\lambda_c=13.89$ 
refers to the critical temperature of the deconfining-preconfining phase 
transition \cite{Hofmann2005}) which rapidly relaxes to $e\equiv\sqrt{8}\pi$ as $T$ increases \cite{Hofmann2007}. 
The right-hand side of Eq.\,(\ref{defF}) is understood as a sum of two contributions obtained by 
constraining the integration to the two sets determined by the following conditions (compare with the 
implementation of the $s$-channel constraint on the four-vertex in \cite{LudescherHofmann2008})   
\begin{align}
  s_{\pm}\left(  \xi,\rho,\lambda, X, f \right) \, \le \, 1 \,,
  \label{supportDimlessCondF}
\end{align}                                                                             
where
\begin{align}
  s_{\pm}\left( \xi,\rho,\lambda, X, f \right) &\equiv
  \left| \frac{f \lambda^{3}}{(2\pi)^2}
  + \frac{\lambda^{3/2}}{\pi}\left(\pm \sqrt{X^2+f} \sqrt{\rho^2+\xi^2+4e^2} - \xi X \right)
        + 4e^2\right|\,,
  \label{supportDimlessF}
\end{align}
and we have defined $f\equiv \frac{F}{T^2}$. These constraints on the loop integration are a specific manifestation of the 
general requirement that in the effective four-vertex 
all three independent 2$\to$2 scattering channels (Mandelstam variables $s$, $t$, and $u$) do not convey momentum transfers larger than 
the scale $|\phi|\equiv\sqrt{\frac{\Lambda^3}{2\pi T}}$. Recall that this scale emerges from an estimate of the thermal ground state based 
on BPS saturated fundamental field configurations \cite{Hofmann2005} which implies that the field $\phi$ is inert. Because 
tree-level massive modes propagate on shell \cite{Hofmann2007} the variables $t$ and $u$ satisfy 
their constraints trivially in our case, and (\ref{supportDimlessCondF}) represents the $s$-channel constraint 
on positive- and negative-frequency loop four-momenta given an external four-momentum on the shell $Y^2-X^2=f$. Taking into account 
conditions (\ref{supportDimlessCondF}), Eq.\,(\ref{defF}) is re-cast as 
\begin{align}
  X^2 - H_F\left( \lambda, X, f\right) = 0 \,,
  \label{NumAlgoF}
\end{align}
where 
\begin{align}
  H_F\left( \lambda, X,f \right) \,  \equiv \, \sum_{\sigma=+,-} \int_{-\infty}^{\infty} \mathrm d\xi\,\int_{0}^{\infty} \mathrm d\rho\,&\,
  \theta\left(1-s_\sigma\left(\xi,\rho,\lambda, X, f \right)\right)\,h_F\left(\xi,\rho, \lambda \right)\,,
  \label{defNumIntHF}
\end{align}
and $\theta(x)$ denotes the Heaviside step function. 

\section{Numerical evaluation and characterization\label{Nec}}

\begin{figure}
\begin{center}
\leavevmode
\leavevmode
\vspace{4.9cm}
\includegraphics{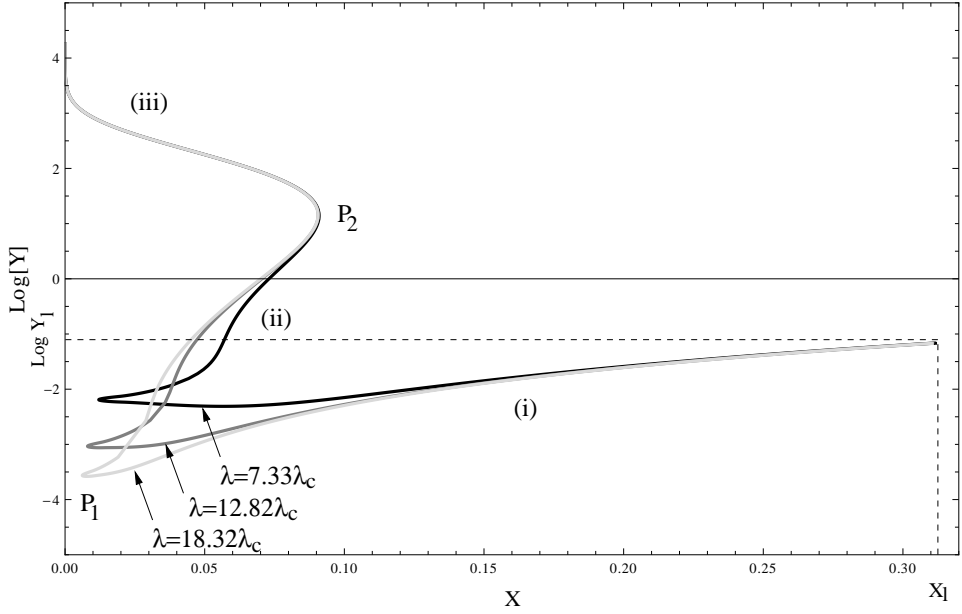}
\end{center}
\caption{\protect{\label{Fig-2}} Three branches of the dispersion for longitudinal modes at 
$\lambda=7.33\,\lambda_c; 12.82\,\lambda_c; 18.32\,\lambda_c$ where $\lambda_c=13.87$. The momenta in all three 
branches (i), (ii), and (iii) are bounded from above by $X_l$ whose high-temperature limit is given as 
$\lim_{\lambda\to\infty}X_l=0.3104=\lim_{\lambda\to\infty}Y_l$. The high-temperature limit of the first point $P_{1}$, where the group velocity $v_g$ becomes singular, $|v_g|\to\infty$, 
is $\lim_{\lambda\to\infty}P_{1,X}=\lim_{\lambda\to\infty}P_{1,Y}=0$. Branch (i), which is supported by 
$P_{1,X}\le X\le X_l$, approaches a photonlike dispersion 
law $Y=X$ as $\lambda\to\infty$. For $P_{2}$ one has $\lim_{\lambda\to\infty}P_{2,X}=0.0906$; $\lim_{\lambda\to\infty}P_{2,Y}=3.188$. Branch (ii), which is supported by 
$P_{1,X}\le X\le P_{2,X}$, exhibits superluminal group velocity, $v_g>1$. Branch (iii), which is supported by $0\le X\le P_{2,X}$, exhibits 
negative group velocity of superluminal modulus.}      
\end{figure}

\begin{figure}
\begin{center}
\leavevmode
\leavevmode
\vspace{4.0cm}
\includegraphics{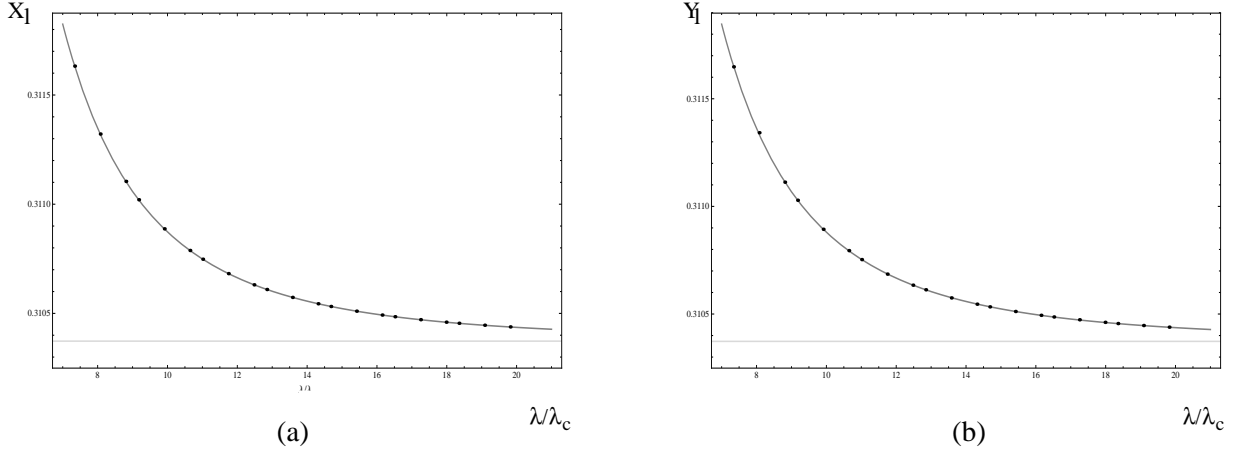}
\end{center}
\caption{\protect{\label{Fig-3}} Plots of the $\lambda$ dependences: (a) $X_l$ and (b) $Y_l$. Dots are the results of the computation based on Eq.\,(\ref{NumAlgoF}), 
and solid lines represent 
the affine-power-law fits of Eq.\,(\ref{powerlawfits}).}      
\end{figure}

\begin{figure}
\begin{center}
\leavevmode
\leavevmode
\vspace{9.5cm}
\includegraphics{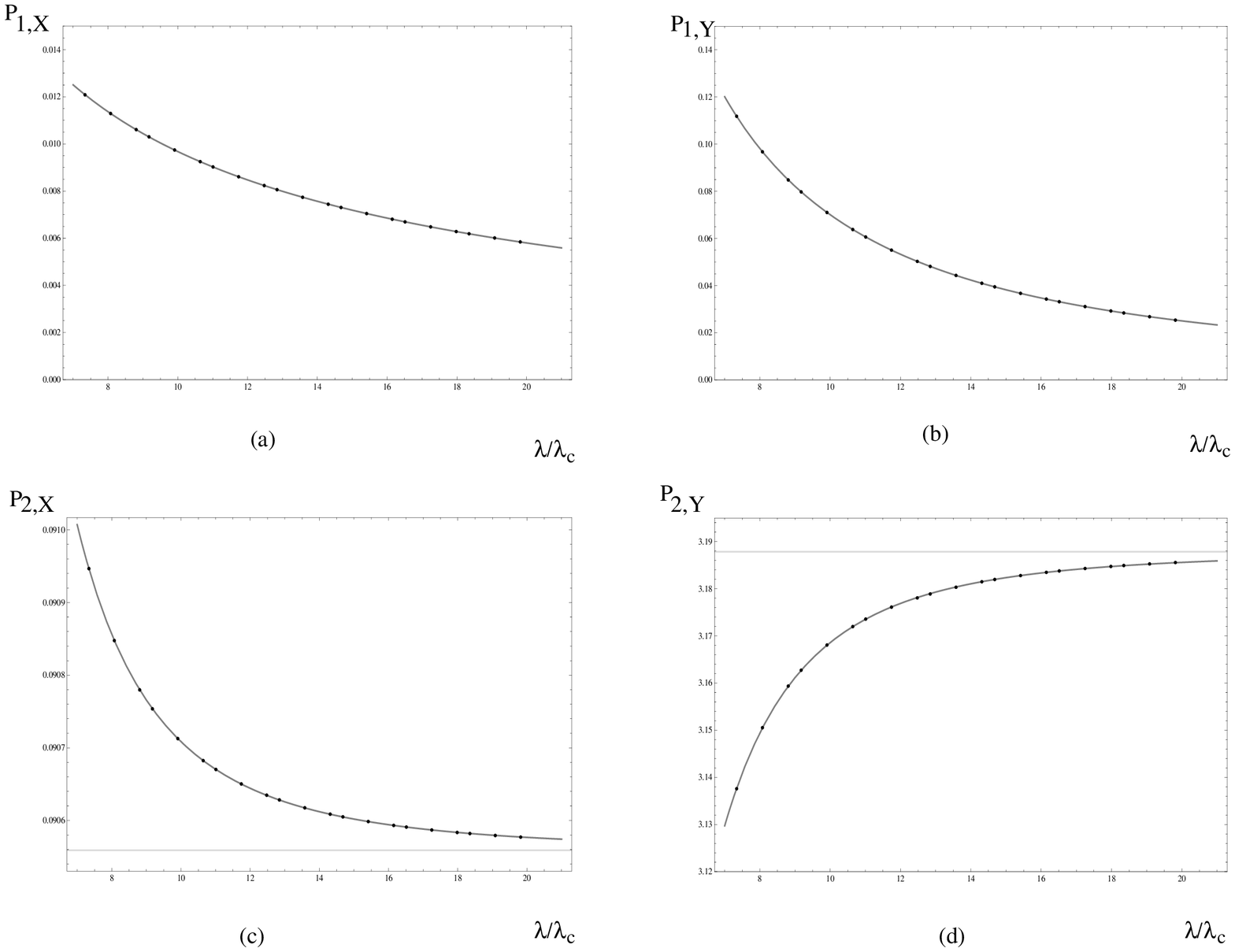}
\end{center}
\caption{\protect{\label{Fig-4}} Plots of the $\lambda$ dependences: (a) $P_{1,X}$, (b) $P_{1,Y}$, (c) $P_{2,X}$, and (d) $P_{2,Y}$. 
Dots are the results of the computation based on Eq.\,(\ref{NumAlgoF}), and solid lines represent 
the affine-power-law fits of Eq.\,(\ref{powerlawfits}). }      
\end{figure}

Let us now solve the gap equation (\ref{NumAlgoF}). Given a value of $\lambda\ge\lambda_c$ and depending on 
$X$, Eq.\,(\ref{NumAlgoF}) 
possesses in $f$ no solution at all or up to three solutions. 
That is, in contrast to the transverse case, where the screening function $g\equiv\frac{G}{T^2}$ is a {\sl function} of $X$ 
at a given value of $\lambda$, there are three branches in the longitudinal case: $f$ is split into several branches which are separated 
by isolated singularities of the group velocity 
$v_g\equiv \frac{d\omega}{d|\vec{p}|}=\frac{dY}{dX}=\frac{d\sqrt{X^2+f}}{dX}$. 
To cope with this situation numerically, we prescribe values for $f$ and $\lambda$ and search the 
root of Eq.\,(\ref{NumAlgoF}) in variable $X$. This root turns out to be unique, 
that is, $X$ is a function of $f$. In Fig.\,\ref{Fig-2} the three branches of 
longitudinal dispersion, $Y=\sqrt{X^2+f}$, are plotted for $\lambda=7.33\,\lambda_c; 12.82\,\lambda_c; 18.32\,\lambda_c$. 
For $1.83\,\lambda_c\le\lambda\le 20.15\,\lambda_c$ excellent fits of the $\lambda$ dependences of $X_l$ and 
the characteristic points $P_1$ and $P_2$ (see Fig.\,\ref{Fig-2}) to affine power laws yield
\eab
\label{powerlawfits}
X_l&=&0.3104+0.4853\left(\frac{\lambda}{\lambda_c}\right)^{-2.987}\,,\ \ \ \ \ \ \ \ \ \ Y_l=0.3104+0.4971\left(\frac{\lambda}{\lambda_c}\right)^{-2.991}\,,\nonumber\\ 
P_{1,X}&=&-0.0006+0.0493\left(\frac{\lambda}{\lambda_c}\right)^{-0.6798}\,,\ \ \ \ \ \ P_{1,Y}=0.0013+2.348\left(\frac{\lambda}{\lambda_c}\right)^{-1.533}\,,\nonumber\\ 
P_{2,X}&=&0.0906+0.1798\left(\frac{\lambda}{\lambda_c}\right)^{-3.080}\,,\ \ \ \ \ \ \ \ \ \  P_{2,Y}=3.188-24.12\left(\frac{\lambda}{\lambda_c}\right)^{-3.098}\,.\nonumber\\ 
\eae
Notice that the exponents $\nu$ in (\ref{powerlawfits}) are quite close to simple fractions: $|\nu|\sim 3,\frac32,\frac23$. 
In Figs.\,\ref{Fig-3} and \ref{Fig-4} plots of $X_l$, $Y_l$ and $P_{1,X},P_{1,Y},P_{2,X},P_{2,Y}$ together with the fitted curves 
are shown, respectively. 
We have checked numerically that besides the powerlike saturation of $X_l$ and $P_{1,X},P_{1,Y},P_{2,X},P_{2,Y}$, 
all intermediate sections of the curves in Fig.\,\ref{Fig-2} saturate rapidly to their limit 
shapes as $\lambda\to\infty$. Finally, we have checked that on branches (i), (ii), and (iii) the contribution of 
diagram A in Fig.\,\ref{Fig-1} to $\Pi_{00}$ vanishes identically. 

Let us now attempt a microscopic interpretation of branches (i), (ii), and (iii). 
The modes in branch (i) are essentially propagating at the speed of light, and it is 
conveivable that the associated fluctuations in the macroscopic electric charge density are 
induced by the slow collective motion of stable magnetic (anti)monopoles \cite{LKGH2008}: Microscopically, 
the acceleration of a given monopole influences the state of motion of adjacent monopoles 
through a radiation field which propagates at the speed of light, and no other spatially 
correlating mechanism exists. This 
is because a given, stable monopole owes its life to the {\sl dissociation}, that is, destruction of a 
large-holonomy (anti)caloron whose services in mediating instantaneous 
spatial correlations from monopole to antimonopole are 
thus not available. On the other hand, the superluminal propagation of radiatively 
induced electric charge density in branches (ii) and (iii) with its larger frequencies $Y$ 
at given value of $X$ should have a microscopic explanation in terms of the instantaneously 
correlated motion of a short-lived monopole and its antimonopole inside a 
small-holonomy (anti)caloron.  

For isotropic and homogeneous thermalization 
the superluminal nature of branches (ii) and 
(iii) does not contradict Special Relativity. This is because the longitudinal modes of 
SU(2)$_{\mbox{\tiny CMB}}$ are, due to a electric-magnetically dual interpretation, propagating {\sl magnetic} fields of 
large wavelength\footnote{For example, $X_l\sim 0.31$ and $T=30\,$K; 3000\,K imply a 
minimal wavelength of about 1.6\,mm; 16$\mu$m.} which do not 
deposit energy in a detector. As a consequence, it is impossible 
to employ these modes for signal transduction.   

\section{High-temperature behavior of energy density\label{htened}}

Here we discuss the high-temperature behavior of the contributions of branches (i), (ii), and (iii) to the total 
thermal energy density $\rho_L$ of propagating longitudinal modes. Moreover, we discuss the 
role of branch (i), which dominates $\rho_L$, in potentially providing magnetic seed fields for MHD dynamos 
upon the breaking of spatial isotropy and homogeneity by astrophysically and large-scale structured matter.   

\subsection{Analytical expressions and numerical results}

In the effective theory for deconfining SU(2) Yang-Mills thermodynamics \cite{Hofmann2005} 
longitudinal modes of the tree-level massless sector are absent on tree level. They emerge as 
thermal quasiparticles by a resummation of the 00 component of the one-loop polarization tensor. 
For the energy density $\rho_L$ of longitudinal 
quasiparticles we have
\eab
\label{rhoquL}
\rho_L(\lambda)&=&\rho_{L,\tiny\mbox{(i)}}(\lambda)+\rho_{L,\tiny\mbox{(ii)}}(\lambda)+\rho_{L,\tiny\mbox{(iii)}}(\lambda)\nonumber\\ 
&=&\frac{T^4}{2\pi^2}
\sum_{B={\tiny\mbox{(i),(ii),(iii)}}}\int_0^{\infty} dX\,X^2\ \nonumber\\  
& &\times\theta(X-X_{B,\tiny\mbox{min}})\,\theta(X_{B,\tiny\mbox{max}}-X)\frac{Y_B(X)}{\exp(Y_B(X))-1}\,,
\eae
where 
\eab
\label{defsintlimits}
X_{{\tiny\mbox{(i)}},\tiny\mbox{min}}&\equiv& P_{1,X}\equiv X_{{\tiny\mbox{(ii)}},\tiny\mbox{min}}\,, \ \ X_{{\tiny\mbox{(i)}},\tiny\mbox{max}}\equiv X_l\,,\nonumber\\ 
X_{{\tiny\mbox{(ii)}},\tiny\mbox{max}}&\equiv& P_{2,X}\equiv X_{{\tiny\mbox{(iii)}},\tiny\mbox{max}}\,, \ \ X_{{\tiny\mbox{(iii)}},\tiny\mbox{min}}\equiv 0\,,
\eae
and $Y_B(X)$ refers to the dispersion law in branch $B=\mbox{(i),(ii),(iii)}$. We have seen in Sec.\,\ref{Nec} that there is a rapid saturation of the curves 
in Fig.\,\ref{Fig-2} to their limit shapes as $\lambda\to\infty$. By virtue of Eq.\,(\ref{rhoquL}) this implies that 
$\frac{\rho_{L,B}(\lambda)}{T^4}$ approach constants at high temperature. One 
has
\eab
\label{rhoquLhighT}
\lim_{\lambda\to\infty}\frac{\rho_{L,\tiny\mbox{(i)}}(\lambda)}{T^4}&=&4.49\times 10^{-4}\,,\ \ \ \ \ \ \ 
\lim_{\lambda\to\infty}\frac{\rho_{L,\tiny\mbox{(ii)}}(\lambda)}{T^4}=7.3\times 10^{-6}\nonumber\\ 
\lim_{\lambda\to\infty}\frac{\rho_{L,\tiny\mbox{(iii)}}(\lambda)}{T^4}&=&2.37\times 10^{-7}\,,\ \ \ \ \ \ \ 
\lim_{\lambda\to\infty}\frac{\rho_{L}(\lambda)}{T^4}=4.57\times 10^{-4}\,.
\eae
Thus, compared to the Stefan-Boltzmann law $\rho_T=\frac{\pi^2}{15}\,T^4\sim\frac23\,T^4$ for two species of transverse, 
massless photons, the correction $\rho_{L}$ is about a tenth of a per mille for $\lambda\gg\lambda_c$. Recall from Sec.\,\ref{Nec} that the presence of 
this energy density in SU(2)$_{\mbox{\tiny CMB}}$ is not measurable radiometrically.  


\subsection{Breaking of isotropy and homogeneity: Magnetic seed fields for galactic MHD dynamos?}

Let us now, on a rather qualitative and speculative level, discuss the potential role of longitudinal modes 
on seeding the generation of magnetic fields in astrophysical structures through the MHD dynamo mechanism. 
According to \cite{Widrow} and references therein the typical strengths of regular magnetic fields are: 2-10\,$\mu$G inside spiral galaxies 
including the Milky Way with coherence-length scales comparable to the scale of the galaxies\footnote{Exceptions range  
up to 50\,$\mu$G  and are correlated with extraordinarily high star-formation rates.}, several $\mu$G in elliptical galaxies 
with coherence-length scales much smaller than those of the spiral galaxies, and 0.1-1\,$\mu$G in galaxy clusters. Finally, from measurements of the 
small-angle CMB anisotropy spectrum a limit of $\sim 10^{-8}$\,G applies to magnetic fields on the present Hubble scale. 

For homogeneous thermalization no regular magnetic field is provided by the longitudinal 
modes of SU(2)$_{\mbox{\tiny CMB}}$ since the plasma is isotropic. (The incoherent contributions of all longitudinal 
modes cancel.) This situation changes upon the introduction 
of anisotropies, e.g. by inhomogeneous thermalization. On one hand, the associated temperature 
gradient would introduce a gradient to the mass of the tree-level massive modes leading to a direction dependence of $\Pi_{00}$ 
and $F$ and thus to the emergence of a nonvanishing magnetic field, compare with Fig.\,\ref{Fig-1}. This field 
could then seed MHD dynamos in astrophysical structures. 
On the other hand, temperature gradients induce a thermoelectric 
effect by virtue of the ground-state physics of SU(2)$_{\mbox{\tiny CMB}}$ \cite{FalquezDA} which could amplify the MHD dynamo mechanism 
by additional current density. 

For an upper bound on the field strength expected from longitudinal modes upon the 
breaking of isotropy and homogeneity in the SU(2)$_{\mbox{\tiny CMB}}$ 
plasma we compute 
\eqb
\label{Best}
\bar{B}\equiv\sqrt{\la\vec{B}^2\ra_{\tiny\mbox{th}}}=\sqrt{2\,\rho_L}=T_c^2\left(\frac{T}{T_c}\right)^2\sqrt{2\times 4.57\times 10^{-4}}\,.
\eqe
Eq.\,(\ref{Best}) represents an upper bound because it assumes 
that the entire energy density of the incoherent longitudinal $B$-field 
modes is, by an external cause, converted into the energy density of a 
coherent $B$-field. Recall that $T_c=2.73\,$K for SU(2)$_{\mbox{\tiny CMB}}$. Use of 
1\,K$=0.862\times 10^{-4}\,$eV and 1\,eV$^2$=14.4\,G yields
\eqb
\label{Bnum}
\bar{B}=2.41\times 10^{-8}\left(\frac{T}{T_c}\right)^2\,\mbox{G}\,.
\eqe
That is, the value of $\bar{B}$ scaled to the present epoch ($T=T_c$) via Eq.\,(\ref{Bnum}) is 
about $\sim 10^{-8}\,$G. Thus the order of magnitude of $\bar{B}$ coincides with the upper bound on Hubble-scale, present-day field strength 
derived from small-angle CMB anisotropies \cite{Widrow}, see above.

\section{Summary\label{sum}}

To summarize, we have computed the dispersion of longitudinal U(1) modes in a 
selfconsistent way at one-loop accuracy in the effective theory for the deconfining phase of SU(2) Yang-Mills thermodynamics. 
These modes do not occur on tree level and represent 
the propagation of radiatively induced electric charge density (charge-density waves).  
We have characterized the three branches of longitudinally propagating electric 
fields which occur at low momenta. Two branches are superluminal. We have discussed why superluminal group velocities do not 
contradict Special Relativity. Namely, the associated long-wavelength and high-frequency modes represent thermalized, propagating {\sl magnetic} 
fields upon an electric-magnetically dual interpretation of SU(2)$_{\mbox{\tiny CMB}}$ 
\cite{Hofmann2005,GiacosaHofmann2005}. Due to their incoherent nature and 
their long wavelengths these modes do not deposit energy in a detector and thus can 
not be employed for signal transduction. As a next step, we have in the high-temperature regime computed the 
energy densities associated with each branch. Their sum represents a correction on the tenth-of-a-per-mille level to 
the Stefan-Boltzmann law for the two polarization states of the massless photon. Finally, we have 
speculated upon the role played by the longitudinal U(1) sector of deconfining 
SU(2)$_{\mbox{\tiny CMB}}$ for the seeding of MHD dynamos to generate the presently 
observed magnetic fields inside astrophysical structures \cite{Widrow}. Coherent 
seed fields can only emerge upon a breakdown of spatial isotropy and homogeneity in the plasma leading to a temperature gradient. 
Driven by gravitational interaction such a gradient is provided naturally by those structures. 
Taking the thermal energy density inherent to the longitudinal, magnetic U(1) sector of 
SU(2)$_{\mbox{\tiny CMB}}$ as an upper bound for coherent field energy, we derive an upper bound for the cosmological present-day field 
strength of about $10^{-8}$\,Gauss. The order of magnitude of this bound coincides with the one obtained from an 
analysis of CMB small-angle anisotropies, see \cite{Widrow} and 
references therein.

\end{document}